\documentclass[aps,12pt]{revtex4}
\usepackage{epsfig}
\newcommand{\be}{\begin{equation}}
\newcommand{\ee}{\end{equation}}
\newcommand{\ba}{\begin{array}}
\newcommand{\ea}{\end{array}}

\def\sqr#1#2{{\vcenter{\hrule height.#2pt
   \hbox{\vrule width.#2pt height#1pt \kern#1pt
      \vrule width.#2pt}
   \hrule height.#2pt}}}
\def\square{{\mathchoice\sqr64\sqr64\sqr{3.0}3\sqr{3.0}3}}
                                                                                
\begin{document}

\title{On the energy momentum dispersion in the lattice regularization}
\author{Bernd A.\ Berg and Zach McDargh}
\affiliation{Department of Physics, Florida State
             University, Tallahassee, FL 32306-4350, USA}
\date{\today} 

\begin{abstract} 
For a free scalar boson field and for U(1) gauge theory finite volume 
(infrared) and other corrections to the energy-momentum dispersion 
in the lattice regularization are investigated calculating energy 
eigenstates from the fall off behavior of two-point correlation 
functions. For small lattices the squared dispersion energy defined
by $E_{\rm dis}^2=E_{\vec{k}}^2-E_0^2-4\sum_{i=1}^{d-1}\sin(k_i/2)^2$ 
is in both cases negative ($d$ is the Euclidean space-time dimension 
and $E_{\vec{k}}$ the energy of momentum $\vec{k}$ eigenstates). 
Observation of $E_{\rm dis}^2=0$ has been an accepted method to 
demonstrate the existence of a massless photon ($E_0=0$) in 4D 
lattice gauge theory, which we supplement here by a study of its 
finite size corrections. A surprise from the lattice regularization 
of the free field is that infrared corrections do {\it not} eliminate 
a difference between the groundstate energy $E_0$ and the mass parameter 
$M$ of the free scalar lattice action. Instead, the relation $E_0=\cosh^{-1}
(1+M^2/2)$ is derived independently of the spatial lattice 
size.
\end{abstract}


\maketitle
\baselineskip = 16pt

{PACS: 11.15.Ha} 

\section{Introduction} \label{sec_Intro}

The propagator of free fields in the lattice regularization suggests
that the continuum Euclidean energy-momentum relation
\begin{eqnarray} \label{Ek}
  \overline{E}_{\vec{\overline{k}}}^2\ =\ \overline{E}_0^2 + 
  \vec{\overline{k}}^2
\end{eqnarray}
becomes replaced by
\begin{eqnarray} \label{EkL}
  E_{\vec{k}}^2\ =\ E_0^2+4\sum_{i=1}^{d-1} \sin(k_i/2)^2\,.
\end{eqnarray}
Here $d$ is the space-time dimension and $E_{\vec{k}}$ the energy 
found from correlation functions of momentum $\vec{k}$ eigenstates. 
Lattice and continuum quantities (the latter here with overlines) are 
related by 
\begin{eqnarray} \label{a}
  \overline{E}_{\vec{\overline{k}}}=\frac{E_{\vec{k}}}{a}
  ~~{\rm and}~~\overline{k}_i=\frac{k_i}{a}\,, 
\end{eqnarray}
where $a$ is the lattice spacing. See, for instance, the textbook by 
Rothe~\cite{rothe}. A divergent correlation length $\xi=M^{-1}$ allows 
for a quantum continuum limit $a\to 0$.

To exhibit violations of (\ref{EkL}) we define a {\it dispersion 
energy} $E_{\rm dis}$ by
\begin{eqnarray} \label{E2dis}
  E_{\rm dis}^2(N)\ =\ E_{\vec{k}}^2(N) - E_0^2(N) -
  4\sum_{i=1}^{d-1} \sin[k_i(N)/2]^2\,,
\end{eqnarray}
where 
$N$ refers to the spatial size of a periodic $N^{d-1}N_d$ lattice. 
Infrared corrections are eliminated by the limit $N/\xi\to\infty$,
where $\xi=M^{-1}$ is a correlation length defined as the inverse
of a suitable mass parameter~$M$.

In the following our equations for the free scalar field hold for 
general $d=1,\,2,\dots,$ while for U(1) we perform Markov chain Monte 
Carlo (MCMC) calculations only in 4D. We estimate energy eigenvalues 
$E_{\vec{k}}$ through the usual cosh-type fits to correlations of 
operators which are in momentum $\vec{k}$ eigenstates
\begin{eqnarray} \label{coshfit}
  C_{\vec{k}}(n_d)\ =\ c_{\vec{k}}\left[e^{-E_{\vec{k}}\,n_d} +
                       e^{-E_{\vec{k}}\,(N_d-n_d)}\right]
\end{eqnarray}
with parameters $c_{\vec{k}}$ and $E_{\vec{k}}$. On our lattices $n_d$ 
takes the values $0,1,\dots,N_d-1$ and the momenta are discretized by
\begin{eqnarray} \label{ki} 
  k_i&=&0,\frac{2\pi}{N},\dots,\frac{2\pi(N-1)}{N}\,,~~i=1,\dots,d-1,
  \\ \label{kd}k_d&=&0,\frac{2\pi}{N_d},\dots,\frac{2\pi(N_d-1)}{N_d}\,.
\end{eqnarray}
We illustrate $\vec{k}\ne 0$ using the lowest non-zero momentum
\begin{eqnarray} \label{k1}
  \vec{k}_1\ =\ (k,0,0)~~{\rm with}~~ k = 2\pi/N
\end{eqnarray}
for which we denote the energy eigenvalue by $E_k=E_{\vec{k}_1}$ and 
the correlation function by $C_k(n_d)$.

For the free scalar field the thus defined energy values do not depend
on $n_d$. In other cases one obtains eigenvalues of the transfer matrix 
only in the $n_d\to\infty$ limit, though the corrections are 
exponentially small in $n_d$. In section~\ref{sec_freefield} we 
derive exact results for the free scalar field. While these calculations 
are straightforward, it seems to us that the difference found between 
the groundstate energy and the mass parameter of the action has not 
been derived in the literature. Amazingly, this difference does not 
disappear in the infinite volume limit $N/\xi\to 0$.

In section~\ref{sec_U1} we turn to U(1) lattice gauge theory (LGT) in 
Wilson's regularization \cite{wilson} and rely on MCMC investigations. 
We set $E^2_0(N)=0$ and use $(E^{\gamma}_k)^2$ in Eq.~(\ref{E2dis}).
Already in the early days of LGT
\begin{eqnarray} \label{Egdis}
  \left(E^{\gamma}_{\rm dis}\right)^2\ =\ 0 
\end{eqnarray}
was estimated in the Coulomb phase for the photon dispersion energy on 
small $4^3\,8$ lattices within the numerical precision available at 
that time \cite{bepa84}. As the direct calculation of the zero-momentum 
photon eigenstate $E^{\gamma}_0$ fails, the estimate via (\ref{Egdis}) 
became a practical method for identifying Coulomb phases in Higgs-type 
models on the lattice. For examples see \cite{lesh86,ka98}.

In Ref.~\cite{makoko} the result (\ref{Egdis}) was again consistent with
the data for simulations in the U(1) Coulomb phase. This time from MCMC 
calculations on much larger $16^4$ lattices. However, in Ref.~\cite{berg10}
\begin{eqnarray} \label{ngdis}
  \left(E^{\gamma}_{\rm dis}\right)^2\ =\ - 0.2102\ (17)
\end{eqnarray}
was reported for a $4^3\,16$ lattice at $\beta=1.1$, where the error 
bar is given in parenthesis and applies to the last digits of the 
estimate (for the Wilson action $\beta>\beta_c$ with $\beta_c=
1.0111331\,(21)$ \cite{arnold} is in the Coulomb phase). In the 
previous work \cite{bepa84,makoko} the finite size effect was
apparently each time swallowed by the statistical error (in 
\cite{makoko} because small systems were not simulated). In 
section~\ref{sec_U1} we present a systematic study of this finite
size behavior for which some of the material is taken from~\cite{mcd}.

A brief summary and conclusions are given in the final 
section~\ref{sec_sum}.

\section{Free scalar field \label{sec_freefield} }

Following the standard approach, e.g.\ \cite{rothe}, the lattice 
regularization for the action of the 4D free scalar boson field 
reads ($n$ and $m$ are integer four vectors)
\begin{eqnarray} \label{SB}
 S_B = -\sum_n\sum_{\mu=1}^4 \phi_n\,\phi_{n+\hat{\mu}} +
        \frac{8+M^2}{2}\, \sum_n \overline{\phi}_n\,\phi_n\,,
\end{eqnarray}
where $\hat{\mu}$ is the unit vector in $\mu$ direction. Next, we 
continue with general dimension $d$ and derive analytical expressions 
for the two-point correlation functions on finite $N^{d-1}\, N_d$ 
lattices with periodic boundary conditions. Subsequently, we compare 
them to the cosh mass fit function (\ref{coshfit}).

For an infinite lattice the two-point correlation function of the 
scalar field $\phi_n$ described by the lattice action (\ref{SB}) 
is derived in \cite{rothe} to be 
\begin{eqnarray} \label{inftyphiphi} 
  \langle\phi_n\phi_m\rangle_{\infty} = 
  \int_{-\pi}^{+\pi} \frac{d^dk'}{(2\pi)^d} 
  \frac{\exp[i\,k'\cdot(n-m)]}{M^2+4\sum_{\mu=1}^d\sin^2(k'_{\mu}/2)}\,,
\end{eqnarray}
where the integration is over each of the $d$ vector components ($d=4$ in
\cite{rothe}). Here we are
interested in the lattice size corrections to this equation. For this 
purpose we replace the integral representation of the Kronecker delta
\begin{eqnarray} \label{delint} \delta_{n\,m} &=& 
  \int_{-\pi}^{+\pi} \frac{d^dk'}{(2\pi)^d}\,\exp[i\,k'\cdot(n-m)]
\end{eqnarray}
used in \cite{rothe} by the one for a $N^{d-1}N_d$ lattice
\begin{eqnarray} \label{del} \delta_{n\,m} &=& 
 \frac{1}{N^{d-1} N_d} \sum_{k'} \exp[i\,k'\cdot(n-m)]\,.
\end{eqnarray}
Here the summation vectors are given by Eq.~(\ref{ki}) and (\ref{kd}).
To match (\ref{delint}) in the limit $N\to\infty$, $N_d\to\infty$ each 
$k_{\mu}$ value can be shifted by $-\pi$. Following the logic of 
\cite{rothe} the two-point correlation function becomes 
\begin{eqnarray} \label{phiphi} 
  \langle\phi_n\phi_m\rangle = \frac{1}{N^{d-1} N_d} \sum_{k'}
  \frac{\exp[i\,k'\cdot(n-m)]}{M^2+4\sum_{\mu=1}^d\sin^2(k'_{\mu}/2)}\,.
\end{eqnarray}
In this paper we are interested in correlation between operators of 
a definite spacelike momentum $\vec{k}=(k_1,\dots,k_{d-1})$ for which 
one expects the functional form (\ref{coshfit}). We define these 
operators by
\begin{eqnarray} \label{phik} 
  \phi_{\vec{k},n_d}\ =\ a_{\vec{k}} \sum_{\vec{n}} 
  \exp\left(-i\,\vec{k}\cdot\vec{n}\right)\,\phi_{\vec{n},n_d}  
\end{eqnarray}
where $a_{\vec{k}}$ is a normalization constant. We want to calculate 
the correlation function
\begin{eqnarray} \nonumber 
  \langle\phi_{\vec{k},n_d} \phi^*_{\vec{k},m_d}\rangle\ \sim\
  \sum_{\vec{n}}\sum_{\vec{m}} \exp\left[-i\,\vec{k}\cdot
  (\vec{n}-\vec{m})\right]\,\langle\phi_n\phi_m\rangle\,.
\end{eqnarray}
Using translation invariance on a periodic lattice this equation
simplifies considerably. First, we note that
\begin{eqnarray} \nonumber 
  \sum_{\vec{n}} \exp\left[-i\,\vec{k}\cdot 
     (\vec{n}-\vec{m})\right]\,\langle\phi_n\phi_m\rangle 
  = \nonumber \sum_{\vec{n}} \exp\left[-i\,\vec{k}\cdot\vec{n}
     \right]\,\langle\phi_{\vec{n},n_d}\phi_{\vec{0},m_d}\rangle\
\end{eqnarray}
holds for all $\vec{m}$. Besides we have
\begin{eqnarray} \nonumber 
     \langle\phi_{\vec{n},n_d}\phi_{\vec{0},m_d}\rangle\ =\
     \langle\phi_{\vec{n},n_d-m_d}\phi_{\vec{0},0}\rangle
\end{eqnarray}
so that it is sufficient to consider
\begin{eqnarray} \nonumber 
  \nonumber \sum_{\vec{n}} \exp\left[-i\,\vec{k}\cdot\vec{n}
     \right]\,\langle\phi_{\vec{n},n_d}\phi_{\vec{0},0}\rangle\
\end{eqnarray}
for which we carry out the $\vec{n}$ and then $\vec{k}'$ summations 
to obtain
\begin{eqnarray} \label{2cor} 
  C_{\vec{k}}(n_d)\ =\ \langle\phi_{\vec{k},n_d} \phi^*_{\vec{k},0}
  \rangle\ =\ \frac{b_{\vec{k}}}{N_d}\sum_{k'_d}\frac{\exp(i\,k'_d\,
  n_d)} {M^2+4\sum_{i=1}^{d-1}\sin^2(k_i/2)+4\sin^2(k'_d/2)}\,.
\end{eqnarray}
The normalization constant $b_{\vec{k}}$ can be chosen to ensure
$C_{\vec{k}}(0)=1$, which fixes also $a_{\vec{k}}$ in (\ref{phik}).
From (\ref{2cor}) it is obvious that the momentum $\vec{k}$ correlation
function $C_{\vec{k}}(n_d)$ simply agrees with the momentum zero 
correlation function $C_0(n_d)$ at the effective mass
\begin{eqnarray} \label{Meff} 
  M^2_{\rm eff}\ =\ M^2 + 4\sum_{i=1}^{d-1}\sin^2(k_i/2)\,.
\end{eqnarray}
It is for the free field sufficient to investigate $C_0(n_d)$ as 
function of its mass parameter $M$. Using neighboring distances 
for the correlation function arguments and the functional form 
(\ref{coshfit}), values for $E_0$ are determined by the equation
\begin{eqnarray} \label{E0} 
  \frac{C_0(n_d)}{C_0(n_d+1)} = 
  \frac{\exp[-E_0\,n_d]+\exp[-E_0\,(N_d-n_d)]}
  {\exp[-E_0\,(n_d+1)]+\exp[-E_0\,(N_d-n_d-1)]}\,.
\end{eqnarray}
Inserting (\ref{2cor}) for $C_0(n_d)$ and $C_0(n_d+1)$, the values for 
$E_0$ turn by numerical inspection out to be independent of $n_d$ and 
$N_d$ for $N_d=2,\,3,\,4,\,\dots$ and $n_d=0,\,\dots\,N_d-1$ as one 
may expect for a massive free scalar field. In particular this includes 
$N_d=2$ and $n_d=0$ for which we find
\begin{eqnarray} \label{cE0} 
  \frac{C_0(0)}{C_0(1)}=\frac{M^{-2}+(M^2+4)^{-1}}{M^{-2}-(M^2+4)^{-1}}
  = \frac{1+\exp(-2\,E_0)}{2\,\exp(-E_0)}\,.
\end{eqnarray}
With a little algebra this simplifies to 
\begin{eqnarray} \label{sE0} 
  \frac{C_0(0)}{C_0(1)}\ =\ 1 + \frac{1}{2} M^2\ =\ 
  \cosh\left(E_0\right)\ =\ 1 + \frac{1}{2}\left(E_0\right)^2
         + \frac{1}{4!}\left(E_0\right)^4 + \dots \,.
\end{eqnarray}
Therefore, $(M-E_0)/M\to 0$ holds in the quantum continuum limit 
$M\to 0$ independently of the spatial lattice size $N$. Correspondingly 
we have
\begin{eqnarray} \label{sEk} 
  \frac{C_{\vec{k}}(0)}{C_{\vec{k}}(1)}\ =\ 1 + 
  \frac{1}{2} M_{\rm eff}^2\ =\ \cosh\left(E_{\vec{k}}\right)\ 
  =\ 1 + \frac{1}{2} \left(E_{\vec{k}}\right)^2 
       + \frac{1}{4!}\left(E_{\vec{k}}\right)^4 + \dots
\end{eqnarray}
for the energies of momentum $\vec{k}$ eigenstates. The dispersion
energy (\ref{E2dis}) becomes \cite{v1}
\begin{eqnarray} \label{E2dis0}
  E_{\rm dis}^2(N)\ =\ \left[\cosh^{-1}\left(\cosh(E_0) +
  2\sum_{i=1}^{d-1} \sin[k_i(N)/2]^2\right)\right]^2
  - E_0^2 - 4\sum_{i=1}^{d-1} \sin[k_i(N)/2]^2\,.
\end{eqnarray}
Dependence on the spatial lattice size comes entirely through 
$k_i(N)$ as the groundstate energy $E_0$ does not depend on $N$.

\begin{figure}[tb] \begin{center} 
\epsfig{figure=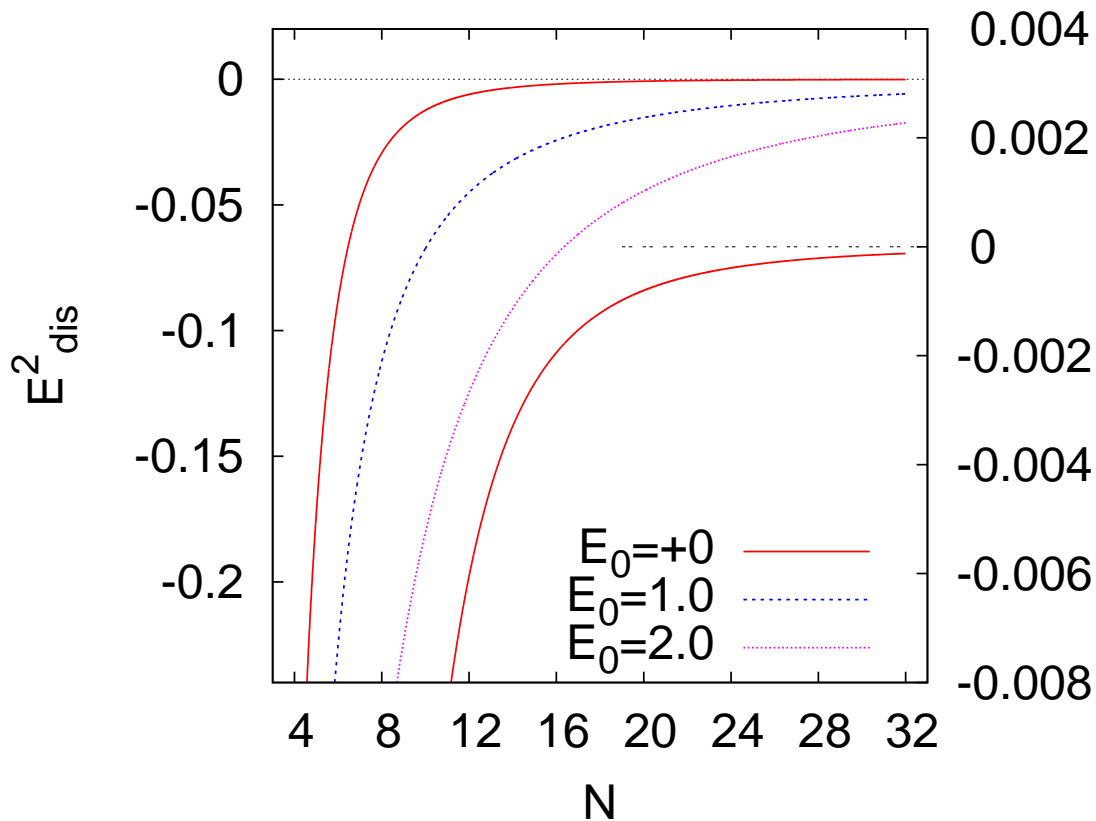,width=\columnwidth} 
\caption{Lattice size dependence of the dispersion energy 
(\ref{E2dis0}) for the free scalar field in 4D. \label{fig_E2dis}} 
\end{center} \end{figure} 

In the Coulomb phase of U(1) lattice gauge theory one expects 
$E_0=0$ and for comparison with the next section we draw in 
Fig.~\ref{fig_E2dis} the function $E^2_{\rm dis}(N)$ for $E_0=+0$, 
which should be interpreted as a very small $E_0>0$ corresponding via 
(\ref{sE0}) to a similarly small $M>0$. These are the first and the 
last of the curves shown, the latter enlarging the approach to zero.
Respectively, the left and the right ordinate apply. Besides $M\to 0$
the continuum energy-momentum relation (\ref{Ek}) requires $k_i(N)
\to 0$ for $N\to\infty$, which follows from (\ref{ki}) and (\ref{kd})
for finite values of the physical momenta $\overline{k}_i$ given
by~(\ref{a}).

With $E_0=1.0$ and $E_0=2.0$ two examples for non-zero groundstate
energies are also given in Fig.~\ref{fig_E2dis}. The left ordinate 
applies. While it can be anticipated that the continuum energy-momentum
relation (\ref{Ek}) becomes only restored for $\xi\to\infty$, we 
expected that the corrections to the lattice energy-momentum relation 
(\ref{EkL}) would decrease with decreasing $\xi/N$, $N$ fixed. The 
curves demonstrate that this is not the case. Though $E^2_{\rm dis}\to 
0$ holds for $N\to\infty$, its finite size corrections increase with 
decreasing correlation length $\xi = M^{-1}$: $\xi=1.04$ for $E_0=1.0$ 
and $\xi=0.567$ for $E_0=2.0$.
The approach to zero relies on $k=2\pi/N\to 0$. If a limit is
considered for which $|\vec{k}|$ stays finite for $N\to\infty$,
the lattice dispersion relation is never fulfilled for energies 
calculated from the two-point correlation functions. This follows
because the lattice size enters into the definition (\ref{Meff}) 
of $M_{\rm eff}$ only through dependence of the momenta on $N$.

\section{U(1) Lattice Gauge Theory} \label{sec_U1}

We consider U(1) LGT with the Wilson action \cite{wilson} on a 4D 
hypercubic lattice with periodic boundary conditions
\begin{eqnarray} \label{U1action}
  S(\{U\}) = \sum_{\square} S_{\square}
\end{eqnarray}
with $S_{\square}={\rm Re}\left(U_{i_1j_1} U_{j_1i_2} U_{i_2j_2} U_{j_2i_1} 
\right)$, where $i_1,\,j_1,\,i_2$ and $j_2$ label the sites circulating 
about the square $\square$ and $U_{ij}$ are complex numbers on the unit 
circle, $U_{ij}=\exp (i\,\phi_{ij})$, $0\le\phi_{ij}<2\pi$, $U_{ji}=
U_{ij}^{-1}$. Expectation values are calculated with respect to the 
partition function 
\begin{eqnarray} \label{U1Z}
  Z = \int \prod_l d\phi_l\,\exp[-\beta\,S(\{U\})]\,,
\end{eqnarray}
where the product is over all links $l=ij$ with $ij$ nearest neighbor
sites of the lattice. Wilson concluded that at strong couplings (small 
$\beta$) the theory confines static test charges due to an area law 
for the product of gauge matrices around closed paths (Wilson loops), 
whereas it describes a massless photon at weak coupling (large $\beta$). 
In between there is a transition between the confined and the Coulomb
phase for which the $\beta$ value has meanwhile been accurately
determined by MCMC simulations, see for instance \cite{arnold}.

However, it turns out that correlations of Wilson loops in 
zero-momentum representations of the cubic group do not give any 
convincing signal for the existence of a massless photon. The 
presumed reason is a noisy power law fall-off, which cannot be 
followed beyond one or two steps in the lattice spacing $a$. A
remedy found in \cite{bepa84} relies on the use of the dispersion
mass (\ref{E2dis}) for the $T_1^{+-}$ representation of the cubic
group. As reviewed in our introduction, until recently estimates 
of $(E^{\gamma}_{\rm dis})^2$ appeared consistent with zero 
\cite{bepa84,makoko}.

\begin{figure}[tb] \begin{center} 
\epsfig{figure=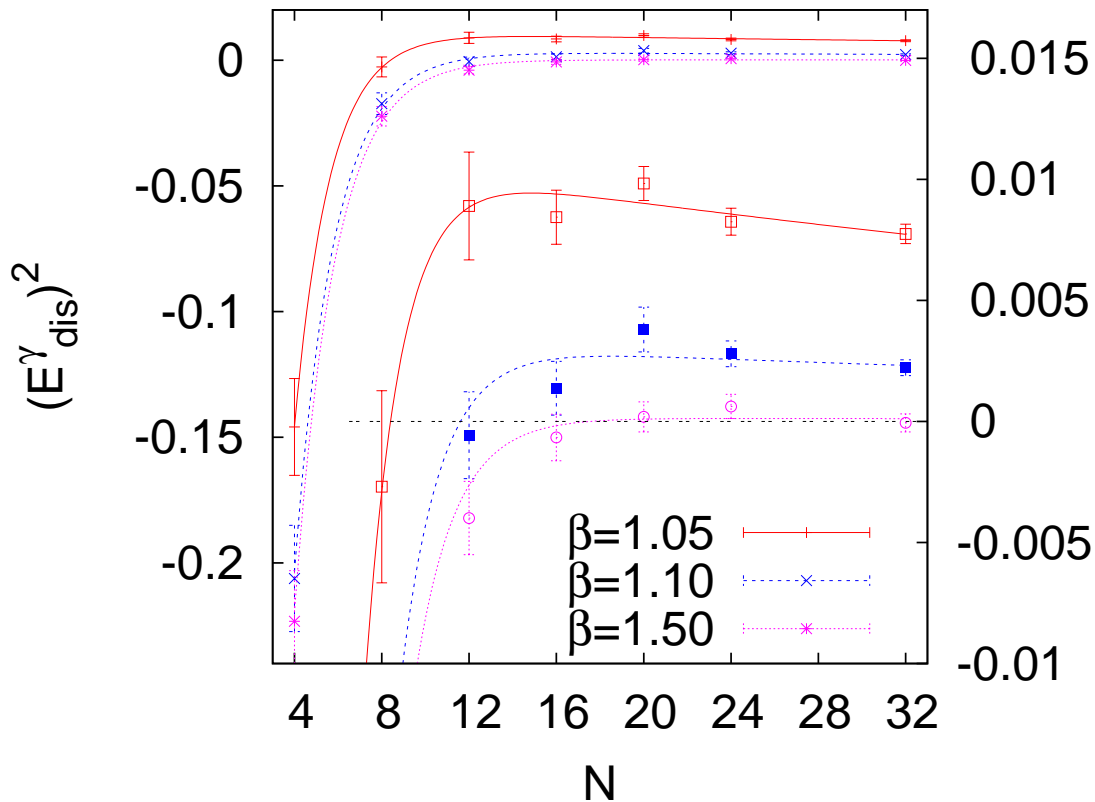,width=\columnwidth} 
\caption{Lattice size dependence of the dispersion energy (\ref{E2dis}) 
for the photon in U(1) LGT. \label{fig_U1exp2}} 
\end{center} \end{figure} 

In view of the deviation (\ref{ngdis}) from zero we decided to follow
the finite lattice size dependence of $(E^{\gamma}_{\rm dis})^2$ from
small $4^3\,16$ to large $32^3\,128$ lattices. Using the 
Metropolis-heatbath algorithm of \cite{babe05} we performed MCMC
calculations for which more details can be found in \cite{mcd} and 
measured correlation functions of the plaquette operator with momentum 
$\vec{k}_1$ (\ref{k1}) and in the $T_1^{+-}$ representation of the 
cubic group. 

\begin{table}[th]
\caption{Estimates of the fit parameters (\ref{U1fits}) (error bars are
 given in parenthesis and apply to the last digits) and the goodness of 
fit $Q$. \label{tab_U1fits}} 
\centering
\begin{tabular}{|c|c|c|c|c|c|} \hline
$\beta$& $a_1$    & $a_2$     & $a_3$      & $a_4$      & $Q$\\ \hline
  1.05 &-1.84 (72)& 0.615 (83)& 0.0117 (19)& 0.0128 (60)&0.49\\ \hline
  1.10 &-1.91 (53)& 0.556 (54)& 0.0035 (19)& 0.0130 (20)&0.31\\ \hline
  1.50 &-1.96 (46)& 0.547 (44)& 0.0002 (16)& 0.010~~(37)&0.89\\ \hline
\end{tabular} \end{table} 

Our results at $\beta=1.05,\,1.1$ and 1.5 are summarized in 
Fig.~\ref{fig_U1exp2}. The left ordinate applies to the upper
three curves. Using in Eq.~(\ref{E2dis}) $E^{\gamma}_0=0$ and
$(E^{\gamma}_k)^2$ as defined after Eq.~(\ref{k1}), they show
$(E^{\gamma}_{\rm dis})^2$ on a scale that includes the estimates
from the smallest $N=4$ lattice. With increasing lattice size an 
approach to zero is observed. The lower three curves, to which the 
right ordinate applies, reveal details. The lines are fits to a 
behavior of the data assumed to be governed by two exponential 
corrections to zero
\begin{eqnarray} \label{U1fits}
  \left(E^{\gamma}_{\rm dis}\right)^2\ =\ 
  a_1\,\exp(-a_2\,N) + a_3\,\exp(-a_4\,N)\,,~~~a_4<a_2\,.
\end{eqnarray}
The fit parameters and the goodness of fit $Q$ are summarized in 
Table~\ref{tab_U1fits}. Large error bars are possible, because the
fit parameter values are correlated with one another.

The enlargements on the the right side reveal for the lower two
$\beta$ values a considerable overshooting of the $(E^{\gamma}_{
\rm dis})^2=0$ value, so that the final approach to zero is from
above. Within our numerical precision it is unclear whether this
overshooting continues for large $\beta$ values. For $\beta=1.5$ 
the coefficient $a_3$ listed in the table is still positive, but 
well consistent with zero.

\section{Summary and Conclusions} \label{sec_sum}

For the free field the mass $M$ of the lattice action (\ref{SB}), 
which enters prominently the lattice propagator (\ref{inftyphiphi})
\begin{eqnarray} \label{propagator} 
  1/\left(M^2+4\sum_{\mu=1}^d\sin^2(k'_{\mu}/2)\right)\,,
\end{eqnarray}
does not agree with the ground state energy, instead (\ref{sE0}) holds 
independently of the lattice size. An astonishing feature of the free 
field exposed in Fig.~\ref{fig_E2dis} is that for fixed lattice size
$N$ the violation (\ref{E2dis0}) of the lattice dispersion relation
increases with decreasing correlation length $\xi$. While such a 
behavior appears natural for the violation of the continuum dispersion
relation, one may have expected that finite size corrections to the 
lattice version decrease generally with decreasing $\xi/N$ and not
just for $\xi$ and momenta $k_i$ fixed. The continuum dispersion 
relation becomes then restored for $k_i\to 0$, which includes the 
quantum continuum limit $E_{\vec{k}} \to 0$ and $k_i\to 0$ with 
$\overline{E}_{\vec{k}}$ and $\overline{k}_i$ of Eq.~(\ref{a}) fixed.

The negative value (\ref{ngdis}) found for the U(1) dispersion energy 
squared is a lattice regularization effect, similarly to the one which 
we have derived with Eq.~(\ref{E2dis0}) analytically for a free scalar 
field. For $E_0=0$ and $k=2\pi/N$ the dispersion energy $E^2_{\rm dis}$ 
(\ref{E2dis}) approaches zero with increasing lattice size for the free 
scalar field as well as for U(1) LGT. A distinction is the overshooting 
of the zero value, which is seen in Fig.~\ref{fig_U1exp2} for U(1) LGT, 
but not in Fig.~\ref{fig_E2dis} for the free field. 
\bigskip

\noindent {\bf Acknowledgements:} This work was in part supported by
the DOE grant DE-FG02-97ER-41022.

\end{document}